\documentclass[a4]{revtex4}

\usepackage{amsmath}
\usepackage{graphicx}
\usepackage{color}

\begin{document}

\title{Single particle nonlocality with completely independent reference states}
\author{J. J. Cooper and J. A. Dunningham}
\affiliation{School of Physics and Astronomy, University of Leeds, Leeds LS2 9JT, United Kingdom}

\begin{abstract}
We describe a scheme to demonstrate the nonlocal properties of a single particle by showing a violation of Bell's inequality.  The scheme is experimentally achievable as the only inputs are number states and mixed states, which serve as references to `keep track of the experiment'.  These reference states are created completely independently of one another and correlated only after all the measurement results have been recorded. This means that any observed nonlocality must solely be due to the single particle state. All the techniques used are equally applicable to massive particles as to photons and as such this scheme could be used to show the nonlocality of atoms.
\end{abstract}

\maketitle

\section{Introduction}

The issue of whether a single particle can exhibit nonlocality has inspired a great deal of discussion and debate in recent years. This controversy has been due, in no small part, to our adherence to the original model for testing locality as prescribed by Einstein, Podolsky, and Rosen (EPR) in their ground-breaking 1935 paper \cite{Einstein1935a}. In the EPR scheme, two separate observers each receive one particle from an entangled pair and measure some property of it. Quantum mechanics predicts that, by comparing their results, they should be able to see a degree of  correlation that surpasses anything possible classically. Einstein, Podolsky and Rosen argued that this strange result provided compelling evidence that quantum mechanics was incomplete and that reality had to be ascribed to  some ``hidden variables"  not contained in the formalism. In 1964, these speculations were put on a firm testable footing when John Bell derived an inequality that had to be obeyed by any local hidden variable theory \cite{bell}. This inequality was tested in a beautiful series of experiments, which showed that violations were indeed possible \cite{Aspect1981, Aspect1982a, Aspect1982b, Tittel1998, Pan2000}. These results led to a reinterpretation of the paradox and the current prevailing view is no longer that quantum mechanics is incomplete, but rather that it cannot be a local realistic theory. That is, if we maintain the view that the physical world is real (i.e. it exists independent of our measurements), then we must abandon locality.

A violation of Bell's inequality in an EPR-type scheme has become the ``industry standard" for confirming the presence of nonlocality. However, such an approach runs into conceptual difficulties when we consider systems consisting of a single particle. If two observers share a single particle state, the detection of the particle at one location precludes the measurement of any property of the particle at the other location. This means that it is not possible to correlate the measurement outcomes and seek violations of Bell's inequality. Of course, if experiments cannot confirm the nonlocality of a single particle, then it is meaningless to attach any reality to it. 

On the other hand,  it might seem odd if nonlocality, which is viewed as a {\it fundamental} feature of the physical world, were not evident for single particles but only appeared as an {\it emergent} property of two or more particles. The question then arises whether the problem of  observing single particle nonlocality is not a fundamental one, but is simply due to the limitations of the schemes used for detecting it.

Tan, Walls and Collett \cite{tan} were the first to propose a different way of detecting nonlocality that could be applied to the single particle case.  Their scheme was later modified by Hardy \cite{hardy} to extend the class of local models it ruled out.  Both these proposals, however, required additional particles to be added to the system, which cast doubt on whether they truly demonstrated a {\it single} particle effect \cite{ghz, vaidman}.
Hardy responded by arguing that the predicted nonlocal properties vanished if the single particle was removed and so any observed nonlocality was directly attributable to this single particle state.
A fertile debate ensued and a number of clever theoretical schemes \cite{peres,revzen, enk, nori} were put forward, however many lacked a clear experimental route to implementation. This left them open to the criticism that they did not represent real testable effects. 

One route that attracted a lot of attention involved  converting single particle entanglement into a more recognizable two-particle entangled state \cite{enk, nori}. This involved passing a single photon through a 50:50 beam splitter and ensuring that each output was incident on a single atom in its ground state. If the photon was carefully tuned to be resonant with an electronic transition of the atom, the result would be an entanglement of two atoms in different states. Such an approach, however, was not without its critics \cite{bjork, drezet}, who argued that it was all an unnecessary matter of semantics since the state could be rewritten in a first quantised picture which does not show entanglement.

Subsequent authors have  proposed \cite{lee} and tested  \cite{babichev} protocols for quantum teleportation  and quantum cryptography \cite{jlee} using only single particle states as the quantum resource.  There is also a scheme  \cite{Dunningham2007a} that makes use of mixed state inputs ensuring that  only experimentally accessible states are required.
A number of experiments have now also demonstrated violations of Bell's inequalities for single particles using a variety of techniques including cavity QED  \cite{moussa} and tomographic detection methods \cite{angelo}.  One such experiment \cite{hessmo, bjork} is based on the setup suggested by TWC and uses a clever combination of beam splitters and polarisers to ensure interferometric stability.  While this experiment gives clear results, it is still subject to the possible `loophole' that the two parties make use of shared reference states, i.e. the reference states have emerged as the two outputs from a single beam splitter. This process is problematic because it could lead to some additional nonlocality being introduced by the reference states. This would make it difficult to ascribe any observed nonlocality to the single particle state.

Evidently, much of the controversy associated with single particle systems stems from this introduction of additional particles. Unfortunately, they seem unavoidable if we are to keep track of the experiment. Instead, we need to ensure that they are added in such a way that they cannot contribute any nonlocality. This means that they must be generated by the two observers using only local operations and classical communication.
In this paper, we show how this can be achieved in a manner that leaves no room for doubt since the two parties each employ a completely independently created mixed state as their reference.
The scheme we present has the further advantage that it is equally applicable to massive particles as to photons and eliminates any need to violate  number conserving superselection rules \cite{hardy,ghz}.

\section{Scheme 1: The Basics}

\begin{figure}[h]
\centering
\includegraphics[width=11cm]{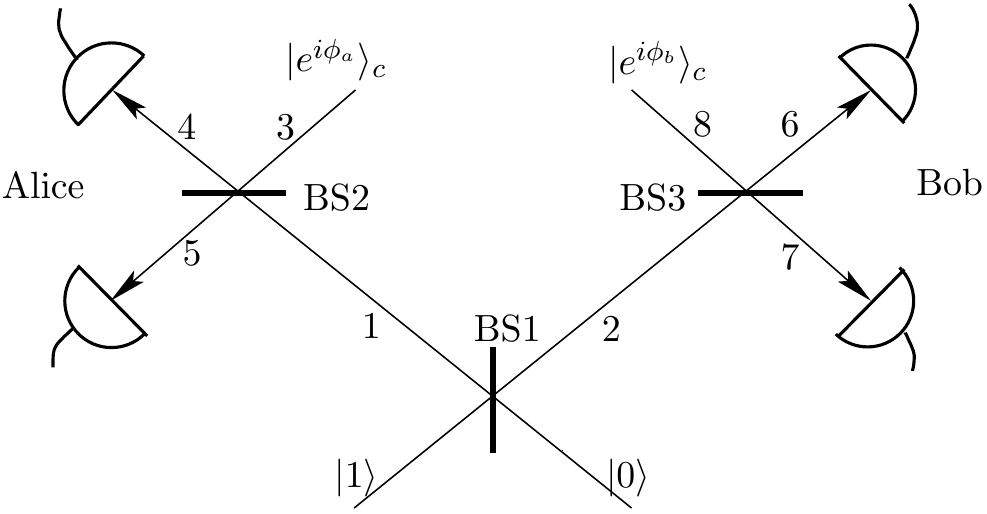}
\caption{Experimental setup for scheme 1 which shows violations of the CHSH inequality for certain values of $\Delta\phi = \phi_{b} - \phi_{a}$.  The reflectivities of BS2 and BS3 can be alternated between $\sin(\xi/2)$ and $\sin(\xi/2+\pi/4)$, and $\sin(\eta/2)$ and $\sin(\eta/2+\pi/4)$ respectively.  The values of $\xi$ and $\eta$ are set so that $\xi - \eta = \pi/4$ to maximise the term $\sin(\xi - \eta) - \cos(\xi - \eta)$ in equation (\ref{s1}).}
\label{scheme1}
\end{figure}

Let us begin by considering the simple scheme shown in Figure \ref{scheme1}.  
For definiteness, we shall describe the case of single photons, however this scheme is equally applicable to massive particles. 
Here we consider a vacuum state, $|0\rangle$, and a single photon state, $|1\rangle$, incident on the the two input ports of a 50:50 beam splitter (BS1).  After this beam splitter the state can be written in a Fock state basis as, 
\begin{equation}
|\psi\rangle = |01\rangle_{12} + i|10\rangle_{12}
\end{equation}
where the first term in each ket represents the number of particles in path 1 and the second, the number in path 2 (represented by the subscripts).  For simplicity, we shall neglect normalisation constants where they are unimportant.  

As shown in Figure~1, path 1 is sent to a second beam splitter (BS2) and path 2 to a third beam splitter (BS3), where they are combined with other states.  The reference state on path 3 is taken to be a coherent state with a mean amplitude of 1 and phase $\phi_a$, i.e. $|e^{i\phi_a}\rangle_c$ and the  reference state on path 8 is taken to be a coherent state with a mean amplitude of 1 and phase $\phi_b$, i.e.  $|e^{i\phi_b}\rangle_c$, where the subscript $c$ is used to distinguish a coherent state from a Fock state.  This means the state of the system before entering BS2 and BS3 can be written as,
\begin{equation}
|\psi\rangle = (|01\rangle_{12} + i|10\rangle_{12})(|0\rangle_3 + e^{i\phi_a}|1\rangle_3+ ...)(|0\rangle_8 + e^{i\phi_b}|1\rangle_8 + ...)
\label{state:BS}
\end{equation}     
where the latter two brackets show the input coherent states expanded in the Fock basis. We can neglect all terms with more than one photon in these coherent states since we are only interested in single photon measurements \footnote{The fact that we consider only certain measurement outcomes may be a source of concern to some readers. However, this is equivalent to Alice and Bob each projecting onto a single-particle state before making their measurements. Since this projection is a local operation for Alice and Bob, it cannot increase the nonlocality in the system and therefore cannot be responsible for any nonlocality that is observed.}.
This state is then transformed by beam splitters BS2 and BS3 and Alice and Bob record which of their detectors at the output ports registers a particle. Afterwards, by comparing their measurement outcomes, they seek nonclassical correlations that are a signature of nonlocality.

\begin{figure}
\centering
\includegraphics[width=15cm]{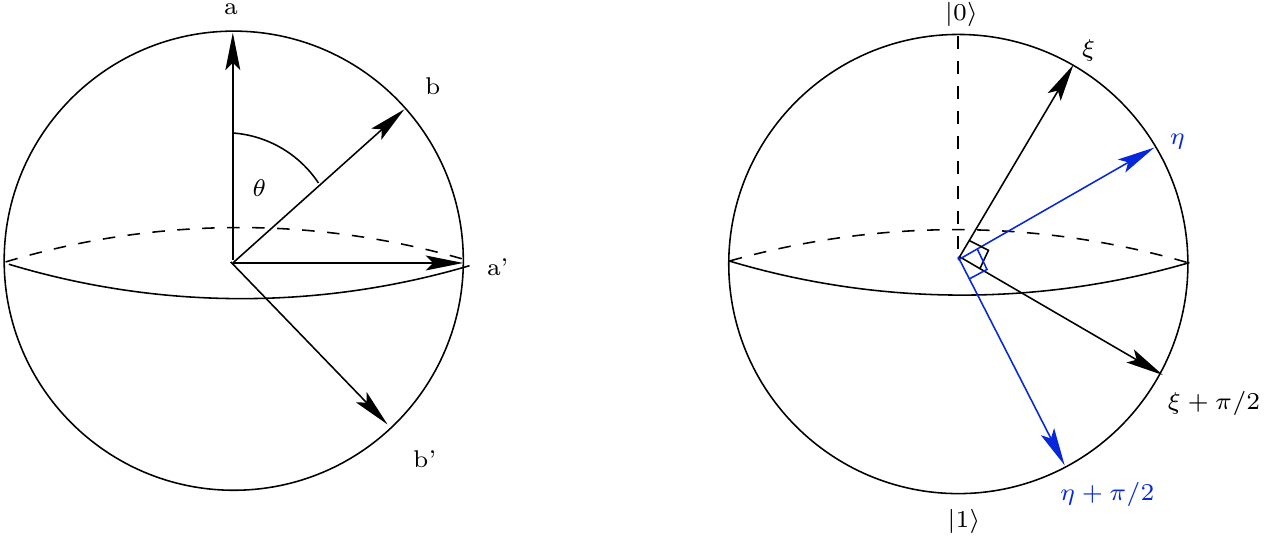}
\caption{Left: When seeking violations of Bell's inequality for EPR pairs of spin-1/2 particles, Alice and Bob each choose two orthogonal spatial directions in which to make measurements. Alice's axes are labelled $a$ and $a'$ and Bob's are labelled $b$ and $b'$. Right: The analogous measurement axes in our scheme for observing single particle entanglement correspond to different directions on the Bloch sphere. In this case, Alice's axes are in directions $\xi$ and $\xi + \pi/2$ and Bob's are in directions $\eta$ and $\eta + \pi/2$. The measurement axes can be adjusted by changing the reflectivities of beam splitters BS2 and BS3.}
\label{bloch}
\end{figure}

Importantly, BS2 and BS3 have variable reflectivities that Alice and Bob can rapidly switch. 
Straightforward schemes exist for implementing beam splitters for atoms \cite{keith1991a, cassettari2000a} with variable reflectivities, e.g. by rapid modulation of the trapping potential \cite{dunningham}.
The role of these beam splitters with variable reflectivities can be understood by considering analogous schemes used to determine the EPR correlations between spin 1/2 particles.  In such schemes, a spin singlet pair is produced: one particle being sent to Alice and the other to Bob.  Alice and Bob can each choose to measure the spin of their particle in one of two spatial directions.  They randomly choose which axis to measure in so as to eliminate the possibility of local hidden variables influencing the results.  
In our single particle scheme, instead of projecting onto different spatial directions, we project onto different superpositions of zero and one particles, i.e. different `directions' when visualised on the Bloch sphere (see Figure~\ref{bloch}).

In analogy, with the spin singlet case, Alice and Bob each choose orthogonal measurement axes and these two sets of orthogonal axes are rotated with respect to each other. In Figure~\ref{bloch}, Alice's axes are shown at angles $\xi$ and $\xi+ \pi/2$ and Bob's are shown at $\eta$ and $\eta + \pi/2$. 
The state represented by $\xi$, for example, is 
\begin{equation}
|\psi\rangle = \cos(\xi/2)|0\rangle + e^{i\phi}\sin(\xi/2)|1\rangle.
\end{equation}
If we take the azimuthal angle, $\phi$ to be $\pi/2$, this is simply a state rotation which can be implemented by a 
beam splitter with reflectivity $\sin(\xi/2)$.
The different detection axes correspond to different reflectivities of BS2 and BS3, e.g. if Alice chooses to measure along axis $\xi$, she picks the reflectivity of BS2 to be $\sin(\xi/2)$ and if she chooses to measure along $\xi+\pi/2$, she picks the reflectivity to be $\sin(\xi/2+\pi/4)$. 

We would now like to see how (\ref{state:BS}) is transformed by the beam splitters.
We are interested only in the cases where Alice and Bob each detect a single particle and we will consider first the case where Alice measures in direction $\xi$ and Bob measures in direction $\eta$.   The transform of BS2 on the possible inputs from paths 1 and 3 is therefore given by,
\begin{eqnarray}
&&|10\rangle_{13} \to \cos(\xi/2)|10\rangle_{45} + i\sin(\xi/2)|01\rangle_{45}\\
&&|01\rangle_{13} \to \cos(\xi/2)|01\rangle_{45} + i\sin(\xi/2)|10\rangle_{45}.
\label{bs}
\end{eqnarray}
Using these transforms one can determine the probability of finding a particle in output paths 4 and 5.  In a similar way (but replacing $\xi$ with $\eta$) one can find the probability of finding a particle in the output paths 6 and 7.  Combining these results it is possible to transform (\ref{state:BS}) to determine the probability of obtaining the state $|1010\rangle_{4567}$ (i.e. the probability of registering a click in path 4, a click in path 6 and nothing in paths 5 and 7). This is given by,
\begin{equation}
P_{4567}(1010) =\frac{1}{2}\left[ \sin^2\left(\frac{\xi + \eta}{2}\right)\cos^2\left(\frac{\Delta\phi + \pi/2}{2}\right) 
+ \sin^2\left(\frac{\xi - \eta}{2}\right)\sin^2\left(\frac{\Delta\phi + \pi/2}{2}\right)\right]
\label{prob1}
\end{equation}
where $\Delta\phi$ is the phase difference between the coherent states, i.e. $\phi_b - \phi_a$.  Using the same method, it is straightforward to show that $P_{4567}(0101) =  P_{4567}(1010)$ and that the probabilties for $|1001\rangle_{4567}$ and $|0110\rangle_{4567}$  are given by, 
\begin{eqnarray}
P_{4567}(1001) &=& P_{4567}(0110) \nonumber  \\
&=& \frac{1}{2}\left[\cos^2\left(\frac{\xi + \eta}{2}\right)\cos^2\left(\frac{\Delta\phi + \pi/2}{2}\right) 
+ \cos^2\left(\frac{\xi - \eta}{2}\right)\sin^2\left(\frac{\Delta\phi + \pi/2}{2}\right)\right].
\label{prob2}
\end{eqnarray}
We can also readily generate the corresponding probabilities for different choices of measurement axes, by simply replacing $\xi$ with $\xi +\pi/2$ in (\ref{prob1}) and (\ref{prob2}), and/or replacing $\eta$ with $\eta +\pi/2$.

We can now use the results in (\ref{prob1}) and (\ref{prob2}) to evaluate the Clauser, Horne, Shimony and Holt (CHSH) version of Bell's inequality \cite{chsh}, which has the form
\begin{equation}
 S\equiv |E(\xi,\eta) + E(\xi+\pi/2,\eta) - E(\xi,\eta+\pi/2) + E(\xi +\pi/2,\eta +\pi/2)| \le 2.
\label{CHSH}
\end{equation}
Any violation of this inequality can be interpreted as evidence of nonlocality in the system. The terms of the form $E(\xi,\eta)$ are expectation values or correlations between Alice and Bob's measurement outcomes given that Alice sets her beam splitter (and hence detection axis) to $\xi$ and Bob sets his to $\eta$.
Suppose on a given run, Alice and Bob each detect their particle in the upper detector, i.e. the detected state is $|1010\rangle_{4567}$. This means their results are perfectly correlated, i.e the correlation is $+1$.
If, on another run Alice detects in the lower detector and Bob in the upper one, i.e. the detected state is $|0110\rangle_{4567}$, then the results are perfectly anticorrelated and so the correlation is $-1$.  It is easy to see that if Alice and Bob each detect only one particle then their results must always be correlated or anticorrelated.

Given that we found the probability of these different cases occurring in (\ref{prob1}) and (\ref{prob2}), we can now find the overall correlation between Alice and Bob's measurement outcomes for settings $\xi$ and $\eta$, i.e. $E(\xi,\eta)$. This is given by,
\begin{eqnarray}
E(\xi,\eta) &=& +\left[ P_{4567}(1010) + P_{4567}(0101)\right] -  \left[ P_{4567}(1001) + P_{4567}(0110)\right] \\
&=& -\sin^{2}\left(\Delta\phi/2+\pi/4\right)\cos(\xi-\eta) - \cos^{2}\left(\Delta\phi/2+\pi/4\right)\cos(\xi+\eta)
\end{eqnarray}
The other correlations can be found in a similar way,
\begin{eqnarray*}
E(\xi,\eta+\pi/2) &=& -\sin^{2}\left(\Delta\phi/2+\pi/4\right)\sin(\xi-\eta) +\cos^{2}\left(\Delta\phi/2+\pi/4\right)\sin(\xi+\eta) \\
E(\xi+\pi/2,\eta) &=& \sin^{2}\left(\Delta\phi/2+\pi/4\right)\sin(\xi-\eta) +\cos^{2}\left(\Delta\phi/2+\pi/4\right)\sin(\xi+\eta) \\
E(\xi+\pi/2,\eta+\pi/2) &=& -\sin^{2}\left(\Delta\phi/2+\pi/4\right)\cos(\xi-\eta) +\cos^{2}\left(\Delta\phi/2+\pi/4\right)\cos(\xi+\eta)
\end{eqnarray*}

Substituting these expressions into (\ref{CHSH}), we obtain
\begin{equation}
S = \left|2\sin^2\left(\frac{\Delta\phi + \pi/2}{2}\right)[\sin(\xi - \eta) - \cos(\xi - \eta)]\right|.
\label{s1}
\end{equation}
We see that this depends only on the relative orientation of Alice and Bob's measurement axes, i.e $\xi - \eta$. This is what we would expect, since schemes that detect EPR correlations for pairs of entangled spin 1/2 particles depend only the relative orientation of Alice and Bob's detectors. 

In order to violate the CHSH inequality, $S$ must have a value greater than 2.  We would now like to find whether there are parameters that give rise to such a violation. $S$ is maximised when each of the two factors in (\ref{s1}) have their maximum value.  This occurs when $\Delta\phi = \pi/2$ and $\xi - \eta = 3\pi/4$. Substituting into (\ref{s1}) gives $S = 2\sqrt2$ which is greater than 2 and as such violates the CHSH inequality. This suggests that Alice and Bob should be able to arrange their detector orientations in such a way that they can observe evidence of single particle nonlocality.  

However, there is a problem in so far as the scheme relies on the two reference coherent states having a fixed relative phase of $\Delta\phi = \pi/2$. If Alice and Bob each prepare their reference states separately, we know that there will be no fixed phase relationship between them. Furthermore, it is well known that the references should be described not as coherent states, but rather as mixed states of the form $ \int_0^{2\pi}|\alpha e^{i\phi}\rangle  \langle \alpha e^{i\phi}|d\phi$ \cite{molmer}.  This avoids problems with violating number conservation superselection rules and means that the two reference states have no absolute phase. If the two references are created separately, their relative phase will vary randomly from shot to shot. This means that on an ensemble average -- which is how these experiments must be performed -- we need to average over all possible phase differences, $\Delta\phi$ in (\ref{s1}).
When we do this, we obtain,
\begin{equation}
S = \left|\sin(\xi - \eta) - \cos(\xi - \eta)\right|,
\end{equation}
which has a maximum value of $S=\sqrt{2}$. Since this is less than $2$, no violation of the CHSH inequality is possible in this case. In the following sections, we show how this problem can be overcome.

\section{Scheme 2: Fixing the Phase}

\begin{figure}
\centering
\includegraphics[width=11cm]{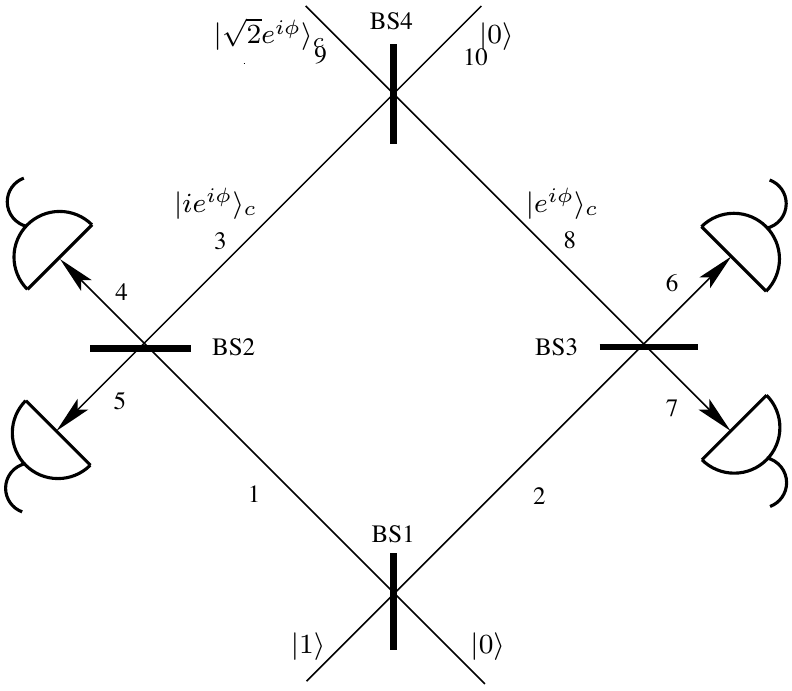}
\caption{Experimental setup for scheme 2.  A violation of the CHSH inequality is achieved on every run of the experiment since the auxiliary state has a fixed phase relationship of $\Delta\phi = \pi/2$ and the reflectivities are chosen such that $\xi - \eta = \pi/4$.}
\label{scheme2}
\end{figure}

The key problem with the scheme outlined in the previous section is that it was not possible to fix the  phase relationship between the two reference states. This problem can be overcome by a simple modification of the set-up (see Figure~\ref{scheme2}).
This time, the two input coherent states in paths 3 and 8 are created by passing a coherent state, $|\alpha\, e^{i\phi}\rangle_c$, and a vacuum state, $|0\rangle$ through a 50:50 beam splitter (BS4).  The output from BS4 is $|i\alpha\, e^{i\phi}/\sqrt2\rangle_{c3}|\alpha\, e^{i\phi}/\sqrt2\rangle_{c8}$ meaning we have a fixed phase relationship of $\pi/2$ between the two input coherent states as required. By picking  $\alpha = \sqrt2$ we also obtain the correct amplitude in each state, i.e. $|ie^{i\phi}\rangle_{c3}$ and $|e^{i\phi}\rangle_{c8}$.  

With these reference states, the total state just before BS2 and BS3 is (c.f. equation~(\ref{state:BS})),
\begin{equation}
|\psi\rangle = (|10\rangle_{12} + i|01\rangle_{12})(|0\rangle_3 + ie^{i\phi}|1\rangle_3)(|0\rangle_8 + e^{i\phi}|1\rangle_8).
\end{equation}
Calculating an expression for $S$ in the same way as above gives,
\begin{equation}
S = 2\left|\sin(\xi - \eta) - \cos(\xi - \eta)\right|.
\label{s2}
\end{equation}
This is exactly as we would expect from (\ref{s1}) since this new scheme fixes the value of the relative phase, $\Delta\phi = \pi/2$.
Consequently the CHSH inequality can be violated on every run of the experiment since this violation depends only on $|\sin(\xi - \eta) - \cos(\xi - \eta)|$, the value of which can be controlled by the experimenters. Furthermore, the input on path 9 can be a mixed state of the form $\rho = \int |\sqrt2\, e^{i\phi}\rangle\langle \sqrt2\, e^{i\phi}|\, d\phi$, since the final result is independent of $\phi$ and so we are free to average over all values of $\phi$ without changing anything.

The coherent reference states, $|ie^{i\phi}\rangle_{c3}$ and $|e^{i\phi}\rangle_{c8}$, are important since they  `keep track of the experiment' and enable Alice and Bob to each make measurements. Since they are separable states, it can be argued that they do not introduce any additional nonlocality into the system and therefore the violation of the CHSH inequality is solely due to the single particle state on paths 1 and 2.
The scheme presented in this section is closely related to the experiment of the Hessmo {\it et al.} \cite{hessmo} except that we have used beam splitters with variable reflectivity instead of polarising beam splitters. This means that our scheme is readily applicable to atoms as well as photons.

\section{Scheme 3: Independent reference states}

Although the scheme presented in the previous section should enable the nonlocality of a single particle to be observed, some readers may feel uneasy about how the reference states were created. Strictly speaking, in order for the reference states to not contribute any of the observed nonlocality, they should be created by only using local operations and classical communication. However, in scheme 2 they were created by a beam splitter, which performs a {\it global} operation on the two states.

For the special case that the input at path 9 is a coherent state (or a mixture of coherent states) and the input at 10 is a vacuum state, then this global operation does not pose any problems since we know that the outputs from BS4 will be completely separable. However, any deviation from this ideal case exposes  the potential `loophole' that some additional nonlocality is introduced by the reference states. This is a particularly pressing concern if we wish to apply this scheme to atoms. Atoms interact with one another through collisions and this leads to nonlinear terms in the Hamiltonian. These nonlinearities tend to number-squeeze the state \cite{greiner, dunningham2}. This means that, in general, the outputs from BS4 will not be separable and, as such, the reference states cannot be eliminated as the source of any nonlocality observed.

A solution to this problem is demonstrated in Figure 3, whereby Alice and Bob each use a completely independent mixed state as their reference. 
Only after all the measurements have been taken are the relative phases of the reference states determined. This eliminates any possibility of nonlocality being due to these states. 
To begin with, we shall consider the case that Alice and Bob each use a mixed state of the form,
\begin{eqnarray}
\rho_{a,b} 	&=& e^{-|\alpha|^2}\sum_{n=0}^\infty\frac{|\alpha|^{2n}}{n!}|n\rangle\langle n|  \nonumber \\
      	&=& \frac{1}{2\pi} \int_0^{2\pi}||\alpha|e^{i\phi_{a,b}}\rangle \langle |\alpha|e^{-i\phi_{a,b}}|d\phi_{a,b},
\label{mixed}
\end{eqnarray}
where the subscripts $a$ and $b$ denote the states belonging to Alice and Bob respectively.  This is a realistic choice for the reference states since it is the natural description of an atomic Bose-Einstein condensate. However, we shall see later that our scheme is not restricted by this choice. Equation (\ref{mixed}) shows that the reference state can be described as a classical mixture of coherent states which are averaged over all phases.  This means that we can decompose the state into coherent states for the purposes of our calculation so long as we correctly average over phases at the end. We stress that these coherent states are just calculational tools: we never physically rely on them.

Referring to Figure~3, we see that Alice combines her state $\rho_a$ with a vacuum state, $|0\rangle$ at BS4.  As discussed above, this input can be decomposed as $|0\rangle_{9}|\alpha\, e^{i\phi_a}\rangle_{c10}$. If the transmittivity of BS4 is chosen to be $1/|\alpha|$, then it allows, on average, one particle to pass through and the output state is therefore $|e^{i\phi_a}\rangle_{c3}|i\sqrt{|\alpha|^2-1}\, e^{i\phi_a}\rangle_{c11}$ \cite{Dunningham2007a}.  Similarly, Bob combines his reference state $\rho_b$ with a vacuum state at a beam splitter (BS5) with transmittivity $1/|\alpha|$. This gives an output $|e^{i\phi_b}\rangle_{c8}|i\sqrt{|\alpha|^2-1}\, e^{i\phi_b}\rangle_{c14}$.  
Comparing with Figure~1, we see that the inputs into BS2 and BS3 are the same as they were in scheme 1 and so the expression for $S$ given by (\ref{s1}) also holds in this case.

The problem, however, is that, just like we found in scheme 1, we do not know the relative phase, $\Delta\phi$ of the reference states. This would suggest that we need to average over all values of $\Delta\phi$, which gives us $S<\sqrt2$, i.e. the CHSH inequality is not violated. 
However, the advantage of the present setup is that we do not need to perform this phase averaging because the crucial phase information is contained in the states on paths 11 and 14. This means that we could make a measurement on these states to determine what the value of $\Delta\phi$ is for each shot of the experiment. Although $\Delta\phi$ is not controllable, it can be determined on each run, which is all we need.
In our scheme $\Delta\phi$ is found simply by combining paths 11 and 14 at a 50:50 beam splitter (BS6) and recording the number of particles, $N_{15}$ and $N_{16}$, detected at ports 15 and 16 respectively.  These are given by,
\begin{eqnarray}
N_{15} &=& (|\alpha|^2-1)\sin^2\left(\frac{\Delta\phi + \pi/2}{2}\right) \\
N_{16} &=& (|\alpha|^2-1)\cos^2\left(\frac{\Delta\phi + \pi/2}{2}\right).
\end{eqnarray}
Combining these, we obtain,
\begin{equation}
\frac{N_{16} - N_{15}}{N_{16}+N_{15}} = 2\cos\left(\Delta\phi + \pi/2\right),
\end{equation}
which allows us to determine $\Delta\phi$ on each run.

In practice, on each experimental run Alice and Bob would record which of their detectors clicked and what setting they had chosen for their beam splitter.  Then a measurement of $\Delta \phi$ would be performed at BS6. Of course, the measurements of $\Delta\phi$ could take place much later -- after all of Alice and Bob's measurements have been completed. The results could then be `binned' into different ranges of $\Delta\phi$ and the correlations $E(\xi,\eta), E(\xi + \pi/2,\eta), E(\xi,\eta+\pi/2)$ and $E(\xi+\pi/2,\eta+\pi/2)$ could be determined for each $\Delta\phi$.  This would enable us to determine the value of $S$ as a function of $\Delta\phi$.  
A plot of how $S$ varies with $\Delta\phi$ is shown in Figure \ref{graph}. We see in particular that in the shaded regions, $S>2$.  This violation of the CHSH inequality would provide an experimental signature of the nonlocality of a single particle.

Finally, we note that this scheme is not restricted to Alice and Bob using reference states of the form of (\ref{mixed}). In particular, if Alice and Bob were to each use number states, $|N_{a}\rangle$ and $|N_{b}\rangle$, the scheme would work in the same way so long as they picked the transmittivities of BS2 and BS3 to be $1/\sqrt{N_{a}}$ and $1/\sqrt{N_{b}}$ respectively. By extension, we see that general mixtures of number states of the form,
\begin{equation}
\rho = \sum_{n=0}^{\infty} P(n)\, |n\rangle \langle n|,
\end{equation}
should also work. The only potential problem with this is that Alice and Bob need to set the transmittivities of their beam splitters according to the number of incident particles. This is to ensure that they skim off a state with a mean amplitude of one.
In practice, so long as the width of the distribution, $P(n)$, is small with respect to the mean number of particles, this should be able to be achieved to a very good approximation.

\section{Conclusions}

We have presented three schemes for observing violations of the CHSH inequality due to a single particle. In general, we have seen that these schemes all require additional reference particles to be added so that Alice and Bob can each make meaningful measurements. The key is that these additional particles are introduced in such a way that they do not contribute any nonlocality to the system.

In the first scheme, Alice and Bob each used an independently created coherent state as their reference. We found that, since there was no fixed relationship between the phases of the reference states, the evidence for nonlocality was washed out and it was not possible to violate the CHSH inequality.
The second scheme resolved this issue by using the two outputs from a single beam splitter as the reference states. This ensured that there was a fixed phase relationship between them. As expected, this allowed for a violation of the CHSH inequality. Indeed this idea is similar to experimental schemes currently used to test single particle nonlocality \cite{hessmo}.
One possible problem with the second scheme, however, is that it relies on a global operation to create the two reference states. This means that, in general, we cannot guarantee that the reference states are completely separable and will not be responsible for the observed violation of the CHSH inequality. 

In the third scheme we overcame this concern by ensuring that Alice and Bob each had independently created mixed states. They then skimmed off part of these states to be used as their reference state and the remaining part was used later to determine the relative phase between their references. 
Although, the relative phase could not be controlled nor predicted in advance, it was knowable on every experimental run. This allows $S$ to be determined as a function of $\Delta\phi$. A theoretical plot is shown in Figure~ \ref{graph}, which predicts that there are values of $\Delta \phi$ where $S>2$, i.e. the CHSH inequality is violated. 
Such a scheme has intriguing prospects for advancing our understanding of the issue of single particle entanglement.
Importantly, is equally applicable to atoms as well as photons as all the employed techniques, such as beam splitters with variable reflectivities, are achievable for both.  

\section{Acknowledgements}

This work was financially supported by a United Kingdom EPSRC Advanced Research Fellowship GR/S99297/01, a University of Leeds Research Fellowship, and a University of Leeds Research Scholarship.

\newpage

%\newpage

\newpage

\newpage

\begin{figure}
\centering
\includegraphics[width=11cm]{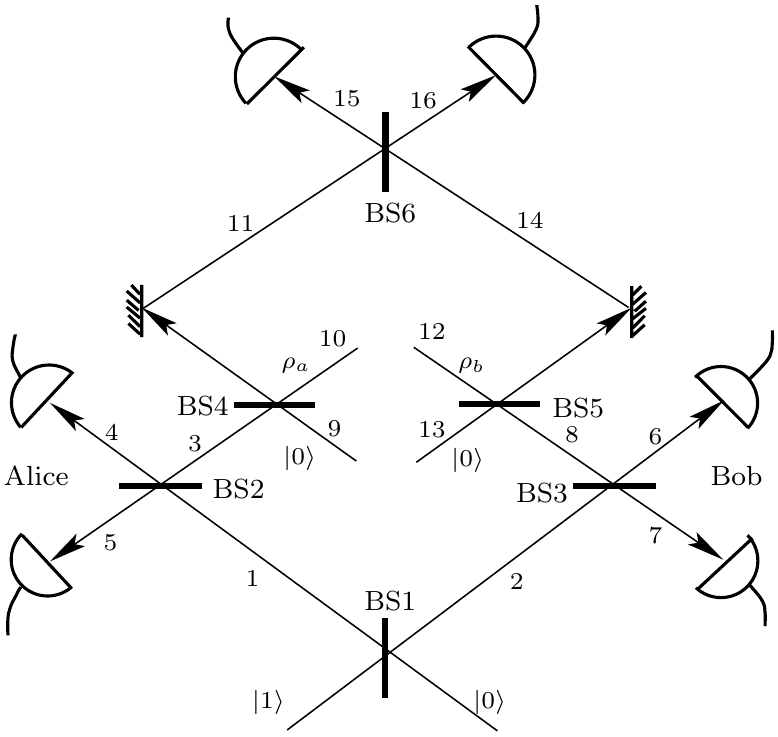}
\caption{Experimental setup of scheme 3 which shows a violation of the CHSH inequality at certain values of $\Delta\phi$.  The clicks at detectors 17 and 18 are used to determine $\Delta\phi$, therefore allowing a value of $S$ to be determined.  Mixed states are used as the auxiliary state and as such no superselection rules are violated.}
\label{scheme3}
\end{figure}

\begin{figure}[h]
\centering
\includegraphics[width=14cm]{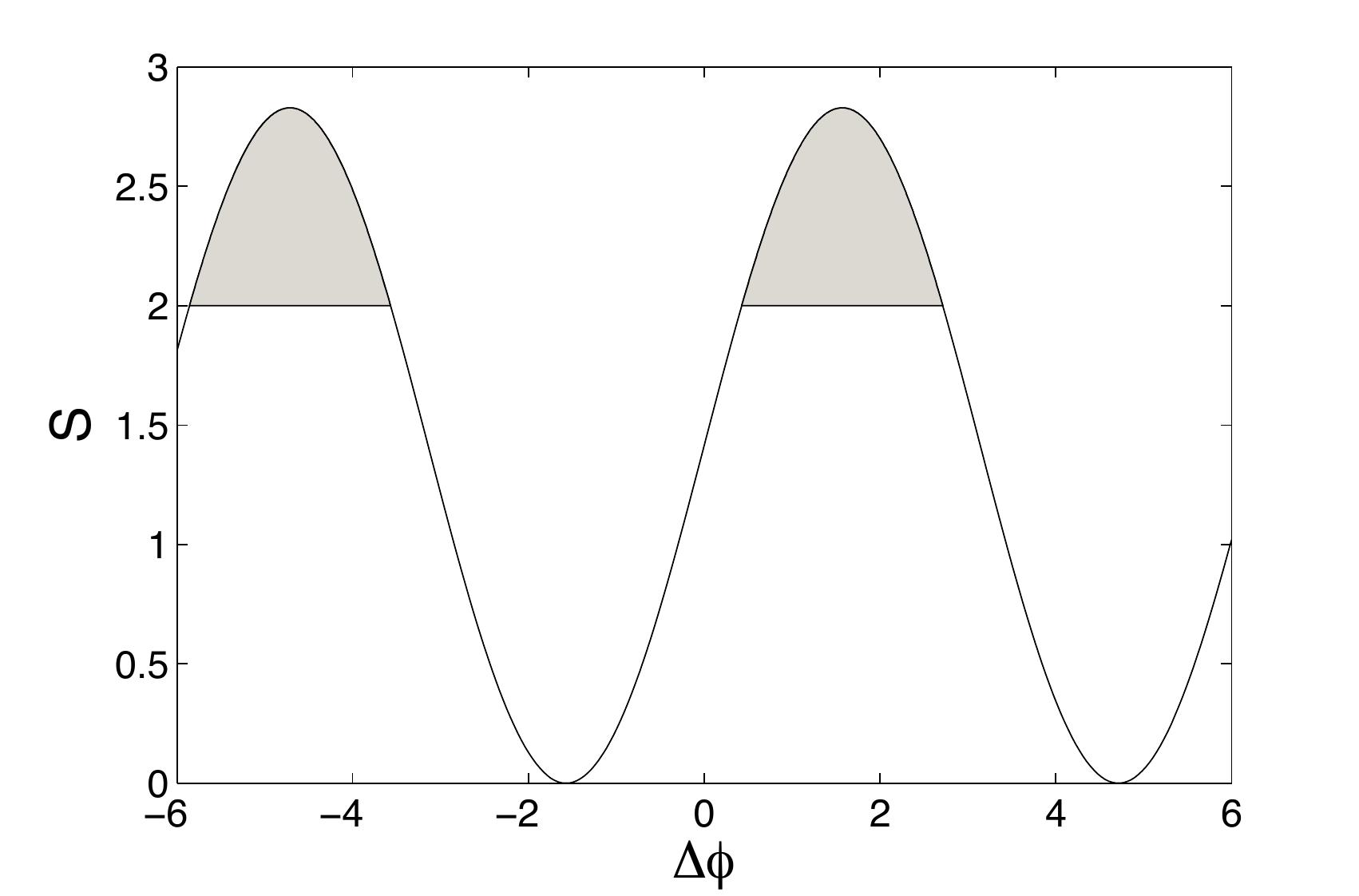}
\caption{This graph shows how $S$ varies with $\Delta\phi$ as given by equation (\ref{s1}).  A violation of the CHSH inequality is evident in the shaded regions where $S>2$.}
\label{graph}
\end{figure}

\end{document}